Title: Recurrent Neural Networks with Long Term Temporal Dependencies in Machine Tool Wear Diagnosis and Prognosis

Corresponding Author: Dr. Binil Starly, Ph.D.

Corresponding Author's Institution:

First Author: Jianlei Zhang

Order of Authors: Jianlei Zhang; Binil Starly, Ph.D.

Abstract: Data-driven approaches to automated machine condition monitoring are gaining popularity due to advancements made in sensing technologies and computing algorithms. This paper proposes the use of a deep learning model, based on Long Short-Term Memory (LSTM) architecture for a recurrent neural network (RNN) which captures long term dependencies for modeling sequential data. In the context of estimating cutting tool wear amounts, this LSTM based RNN approach utilizes a system transition and system observation function based on a minimally intrusive vibration sensor signal located near the workpiece fixtures. By applying an LSTM based RNN, the method helps to avoid building an analytic model for specific tool wear machine degradation, overcoming the assumptions made by Hidden Markov Models, Kalman filter, and Particle filter based approaches. The proposed approach is tested using experiments performed on a milling machine. We have demonstrated one-step and two-step look ahead cutting tool state prediction using online indirect measurements obtained from vibration signals. Additionally, the study also estimates remaining useful life (RUL) of a machine cutting tool insert through generative RNN. The experimental results show that our approach, applying the LSTM to model system observation and transition function is able to outperform the functions modeled with a simple RNN.

**Edward P. Fitts Department of Industrial and Systems Engineering**
**College of Engineering**
**North Carolina State University**
**111 Lampe Drive, Raleigh, NC 27607**

Aug 2nd , 2017

Professor A. Abraham
Editor in chief, Engineering Applications in Artificial Intelligence

Dear Dr. Abraham,

We are pleased to submit our original manuscript entitled "**Recurrent Neural Networks with Long Term Temporal Dependencies in Machine Tool Wear Diagnosis and Prognosis**" for your review and consideration to be published in the Journal of Engineering Applications in Artificial Intelligence.

Machine Cutting Tool Diagnosis and Prognosis has been researched for decades. Yet, many are limited by the complexities of predicting cutting tool wear in a dynamic production based environment. With advancements made in deep learning, our work for the first time, has built on a modified recurrent neural network, specifically using long-short term memory to help diagnose and predict tool wear. In addition, our model also can allow one-step, two-step and even multi-step ahead prediction of tool wear amounts, along with remaining useful life prediction. This work is unique in that, to the best of our knowledge, we have not seen such application of deep learning networks to machine tool research.

All authors have contributed to this article. This work has not been published elsewhere before. All illustrations in the article are also our own.

Please let me know if you need further information regarding this submission.

Thank you for your time and I look forward to hearing from you.

Sincerely yours,

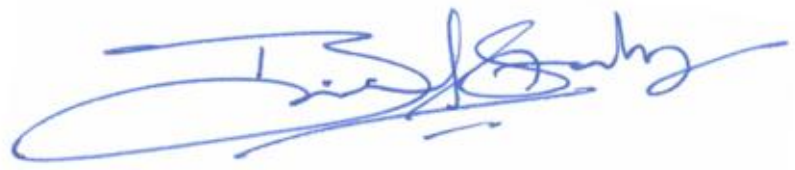

Binil Starly, Ph.D.
Associate. Professor, Data Intensive Manufacturing Laboratory
Tel: 1 919 515 1815; Email: bstarly@ncsu.edu

**Highlights (for review)**

- Recurrent Neural Network with Long Short Term Memory (LSTM) for Predicting Machine Cutting Tool Wear
- One-Step and Two-Step Look Ahead Prediction of Cutting Tool Wear
- Remaining Useful Life Prediction from RNN-LSTM
- Comparison of Simple RNN, LSTM and Gated Recurrent Units(GRU)

# Recurrent Neural Networks with Long Term Temporal Dependencies in Machine Tool Wear Diagnosis and Prognosis

Jianlei Zhang, Binil Starly*
*Edward P. Fitts Department of Industrial and Systems Engineering North Carolina State University, Raleigh, NC 27695 USA. Contact Author: bstarly@ncsu.edu*

**Abstract**
Data-driven approaches to automated machine condition monitoring are gaining popularity due to advancements made in sensing technologies and computing algorithms. This paper proposes the use of a deep learning model, based on Long Short-Term Memory (LSTM) architecture for a recurrent neural network (RNN) which captures long term dependencies for modeling sequential data. In the context of estimating cutting tool wear amounts, this LSTM based RNN approach utilizes a system transition and system observation function based on a minimally intrusive vibration sensor signal located near the workpiece fixtures. By applying an LSTM based RNN, the method helps to avoid building an analytic model for specific tool wear machine degradation, overcoming the assumptions made by Hidden Markov Models, Kalman filter, and Particle filter based approaches. The proposed approach is tested using experiments performed on a milling machine. We have demonstrated one-step and two-step look ahead cutting tool state prediction using online indirect measurements obtained from vibration signals. Additionally, the study also estimates remaining useful life (RUL) of a machine cutting tool insert through generative RNN. The experimental results show that our approach, applying the LSTM to model system observation and transition function is able to outperform the functions modeled with a simple RNN.

## 1. Introduction

Data-driven approaches to predicting machine condition has seen rapid advancements due to low cost sensor technology and reduced computational cost. In machining based operations, a critical asset to obtaining superior part quality are the cutting tool inserts. These inserts continuously wear out and are replaced in a production based environment. The continuous estimation of the tool wear classifications caused by abrasion, erosion or tool breakage is necessary to improve the reliability of the metal-cutting manufacturing processes. Tool breakage can be detected by robust power monitoring techniques, leading to emergency shut-off of the machine. However, tool wear due to abrasion and erosion are hard to estimate in continuous production environments. At present, production environments resort to offline detection through the use of in-machine laser probes which measure wear amounts on the cutting tool surface or a scheduled replacement of cutting tools based on regular machining time intervals. Non-optimized changeover of cutting tool inserts can lead to lost production efficiency and reduced part quality. Detection and prediction of tool wear amounts in real-time is necessary to improve discrete manufacturing systems operations.

Research in diagnosing and prediction of tool wear has been carried out for decades. This problem is particularly difficult due to modeling difficulties involved in a non-linear time-variant process. To cope with this problem, prediction techniques have been applied to estimate future tool wear amounts but often resort to intrusive sensing techniques which are limited in practical industrial implementation. Sensor signals are dependent on a number of factors including machine state, machining conditions and position of the sensor. Adding to the complexity is the apparent variability in the cutting tool, inconsistencies in the material properties of the workpiece and the larger number of machining conditions that must be accounted for.

Several methods have been proposed in the literature for tool wear estimation and prognosis through both linear and non-linear prediction. The most widely used non-linear prediction for tool wear methods involve neural networks (NN). Various works have shown the power of NN in terms of predicting tool wear amounts which can meet accuracy requirements. They do however require large data to collect over periods of time under various machining conditions. Recurrent Neural Networks (RNN) were proposed as an improvement over regular feed-forward neural networks to account for cyclic connections over time. Each successive time-step in the machining states are stored in the internal state of the network to provide a temporal memory. However, conventional RNN did suffer from gradient vanishing or exploding issues (Hochreiter, et al., 2001) when it is trained with gradient based learning and back-propagation through time (BPTT) methods. New variations in RNN, the gated recurrent neural networks, including long short-term memory (LSTM) and Gated Recurrent Units (GRU), have been developed to address this problem. Long Short-Term Memory (LSTM) is a RNN architecture that is excellent at remembering both short-term and long term dependencies. Originally designed by Hochreiter, et al. (1997), this method was developed to address the issue of the vanishing and/or exploding gradient problem when number of layers in the neural network is increased. These RNNs in the form of LSTMs have been very successfully used in speech recognition, language modeling, network traffic regulation, visual recognition and many others.

The objective of this study is to improve tool wear estimation and RUL prediction by utilizing an LSTM based RNN framework, which makes more effective use of model parameters to train system observation and transition models of tool wear processes. We utilize minimally intrusive vibration sensors to collect the signal online and help train the RNN model. We train and compare our LSTM models with other forms of RNN. The solution approach can be extended to assess other critical assets of a manufacturing machine. The organization of this work is as follows: First, we provide a brief overview of related work in estimating tool wear using neural networks. The gated RNN architecture and equations are laid out as the basis of our work. We detail our experimental setup and selected cutting conditions for tool wear diagnosis and prognosis. We finally report on the ability of the method to provide real time estimation, one-step and two-step look ahead predictions of tool wear state, as well as RUL prediction.

## 2. Related Work

Neural networks (NN) were widely used for modeling and predicting tool wear because they can learn complex non-linear patterns without the need for an underlying data relationship. They have strong self-learning and self-adaptive capabilities. Most previous work do not consider the sequential nature of the sensor data or are reliant on a combination of intrusive sensors that limits its practical use. Feature extraction and selection are most commonly performed, which therefore require some form of human labor involved in deciphering the sensor signals.

Wang, et al.'s (2008) work proposed the Fully forward connected neural network (FFCNN), and the optimized FFCNN by pruning some weights among the network. It was trained by the extended Kalman filter (EKF), rather than the back propagation (BP). The result shows that it has better performance than Multilayer Perceptron (MLP) trained with BP. Özel, et al. (2005) applied the feed forward neural network (FFNN) to estimate the surface roughness, and tool flank wear using a variety of cutting conditions. Bayesian regularization with *Levenberg–Marquardt* method is used in training to obtain good generalization capability and help in determining the number of neurons in the hidden layers. However, these two methods still did not consider time dependencies of the inputs and outputs to model, and cannot make prediction of the future tool wear.

Kamarthi, et al. (1995) and Luetzig, et al. (1997) proposed a method composing radial basis function networks (RBF) networks and a RNN. RBFN approach fuses and maps the sensor signal to the tool wear, which leads to the estimation of tool wear amounts. The method considers the time dependencies of the inputs, the sensor signal, and the past estimated tool wear, which is better than the FFNN and MLP methods. However, the number of time steps of temporal dependencies is predefined in the model, rather than determined by any input value. Moreover, the developed method is limited to estimating the tool wear amount online, but cannot predict the future tool wear and hence the RUL of the cutting tool.

Overcoming the limitation of simple RNN, the gated recurrent neural networks were proposed in tool wear monitoring and prediction. The simple RNN can only realize one-step ahead and two-step ahead prediction, but its horizon on prediction is limited, meaning that it is limited in its ability to be extended to RUL prediction. The conventional RNNs cannot capture long term-dependencies, resulting in the NN not able to capture previous long-term history into current estimation. This is due to the well described problem of the vanishing or exploding gradient as the number of time steps in the RNN increases. Therefore, gated recurrent neural networks, including LSTM and GRU, were developed to avoid this problem.

The motivation for this work is that a model-free approach which does not need an underlying analytical model and it would be a better approach to capture long term dependences, prediction in an arbitrary horizon and remaining useful life prediction for a non-linear time variant process. The advantages of our approach over existing ANN and RNN approaches are:

- LSTM based RNNs capture time dependencies and make prediction over conventional NNs that are based on Feed Forward Neural Networks.
- The proposed approach does not have the limitation of limited prediction horizon period, which is the major drawback of time-delayed neural networks.
- Time delayed based RNNs, although can achieve time dependencies, the method cannot realize arbitrary time step-ahead and RUL prediction.
- Our approach of including a model System Transition and System Observation function separately can naturally realize diagnosis and prognosis within the LSTM based RNN architecture.

## 3. Construction of an RNN State Space Model for Tool Wear Monitoring

The proposed approach is a model-free method based on Recurrent Neural Networks by formulating a temporal dependent relationship between in-process indirect measurement and in-process real machine state, and the relationships between the real machine states themselves in the temporal domain. Due to its model-free nature, RNN can model the non-linear relationship between their inputs and outputs. Therefore, the assumption pertaining to the Markov model and linearity can be avoided. Long term dependencies cannot be modeled by conventional RNN approaches. Here we adapt the Long Short-Term Memory (LSTM) and Grated Recurrent Units (GRU) to model long term dependencies for multi-step prognosis of tool wear processes. We build a system observation function and a system transition function with simple RNN, LSTM and GRU. We split experimental data collected to a training set, validation and a test set. We train the system observation function and the system transition function separately to help tune its individual parameters and to implement independent training protocols. More importantly, the independent training also allows us these functions to go through varying training iterations. A comparison will be made between RNN, LSTM and GRU for the system observation and system transition function respectively, and then the best RNN model will be selected for one of these functions. With the trained models, an online machine state estimation is realized by the indirect sensor measurement. Building on the idea of generative models (Graves, 2013), an arbitrary time step ahead prediction can be updated in real-time, as well as to predict Remaining Useful Life (RUL).

### 3.1. System Transition Function

For dynamic mechanical systems, the system state may have temporal dependencies. More specifically, the current system state is affected by its past system states. Therefore, the System Transition function can be generically written as follows:

$y_{t+1} \sim p(y_{t+1}|y^t; \theta)$, where $y^t = (y_1, y_2, \ldots, y_t)$

Here, $y_t$ is the direct measurement of the system state. When the system transition function is modeled as an RNN, the function can be rewritten as:

$y_{t+1} \sim p(y_{t+1}|y_t, h_t^{trans}; \theta)$

where $h_t^{trans}$ is the hidden nodes for the system *transition* function, with superscript letter '*trans*' denoting it. The current hidden node $h_t^{trans}$ is a function of the past hidden node $h_{t-1}^{trans}$ and current input $y_t$. so via the hidden node, current hidden node bear all the current and past information to the inputs.

$h_t^{trans} = f(h_{t-1}^{trans}, y_t; \theta)$

so that $h_t^{trans} = g_t^{trans}(y_1, y_2, \ldots, y_t; \theta)$,

which the function $g_t^{trans}$ takes the whole pass sequence $(y_1, y_2, \ldots, y_t)$, so that the system state at time t+1 can be modeled given all the past system state information. Note that in our implementation, the variable here is a continuous value and its distribution is assumed as unimodal, although it is possible to extend to the arbitrary value with arbitrary distribution.

### 3.2. System Observation Function

The direct measurement of the real-system state is often not practically feasible to be collected in production environments, such as actual tool wear amounts on cutting tool inserts. The system state can be inferred from indirect measurements, such as one or more sensor signals. Also, in most cases, there is not a good function that can build a relationship between the direct measurement of the system state and its associated indirect measurement. Also, any developed relationship must consider temporal effects. Specifically, with availability of any past direct measurements of the system state, the determination of current system state should refer to all past indirect measurements. Therefore, the system observation function can be written as:

$y_t \sim p(y_t|x^t, \theta)$, where $x^t = (x_1, x_2, \ldots, x_t)$

here, $x_t$ is the indirect measurement at time step t; and $y_t$ is the direct measurement of the system state.

With RNN, the system observation function can be rewritten as:

$y_t \sim p(y_t|x_t, h_t^{obs}, \theta)$
where $h_t^{obs}$ is the hidden nodes for the observation function with superscript letter '*obs*' denoting it.

Similar to $h_t^{trans}$ , the hidden nodes for the system *transition* function,
$h_t^{obs} = f(h_{t-1}^{obs}, x_t; \theta)$
so that $h_t^{obs} = g_t^{obs}(x_1, x_2, \ldots, x_t; \theta)$,
which the function $g_t^{obs}$ takes the whole pass sequence $(x_1, x_2, \ldots, x_t)$.

### 3.3. Long Short Term Memory (LSTM) Cell

LSTM cell is an advanced version of the RNN cell, which are excellent at remembering values for either long or short time periods. It consists of repeating memory cell units, each of which consists of interacting elements – the cell state, and three gates that protect the cell state - input gate, a forget gate and an output gate (Figure 1). The gates, via the activation function, control the flow of information to update the cell state. The input gate controls the new input flow into the memory, and the output gate controls the extent of the flow from memory into the output nodes. The forget gates, which implements in the self-loops, control the gradient flow from previous cells from the long temporal past, thereby adaptively forgetting or including the distant memory. Additionally, with gated self-loops controlled by the weights, the time scale of integration can also be controlled by the forget gates (Goodfellow, et al. 2016). At each time step t, the hidden state, $h_t$ is updated by $x_t$, the current data at the time step, the hidden state at previous time $h_{t-1}$, input gate i, forget gate f, output gate o and the internal memory cell c. U and W are the weight matrices.

input gates:
$i = \sigma_i(x_t U^i + h_{t-1} W^i)$

forget gates:
$f = \sigma_f(x_t U^f + h_{t-1} W^f)$

output gates:
$o = \sigma_o(x_t U^o + h_{t-1} W^o)$
$g = tanh(x_t U^g + h_{t-1} W^g)$, where g is candidate hidden state which is used to calculate the real hidden state.
$c_t = c_{t-1} \cdot f + g \cdot i$
$h_t = tanh(c_t) \cdot o$

As seen from the above equations, all the gates can be determined by the input $x_t$ and hidden nodes $h_{t-1}$, with weights that can be tuned during the training process.

Finally, the output calculation, the output $y_t$ is written as
$y_t = \sigma_y(h_t W^y + b_y)$
where, σ is activation function, $x_t$ is the input at time step t; and $y_t$ is the output at time step t. and b is bias.

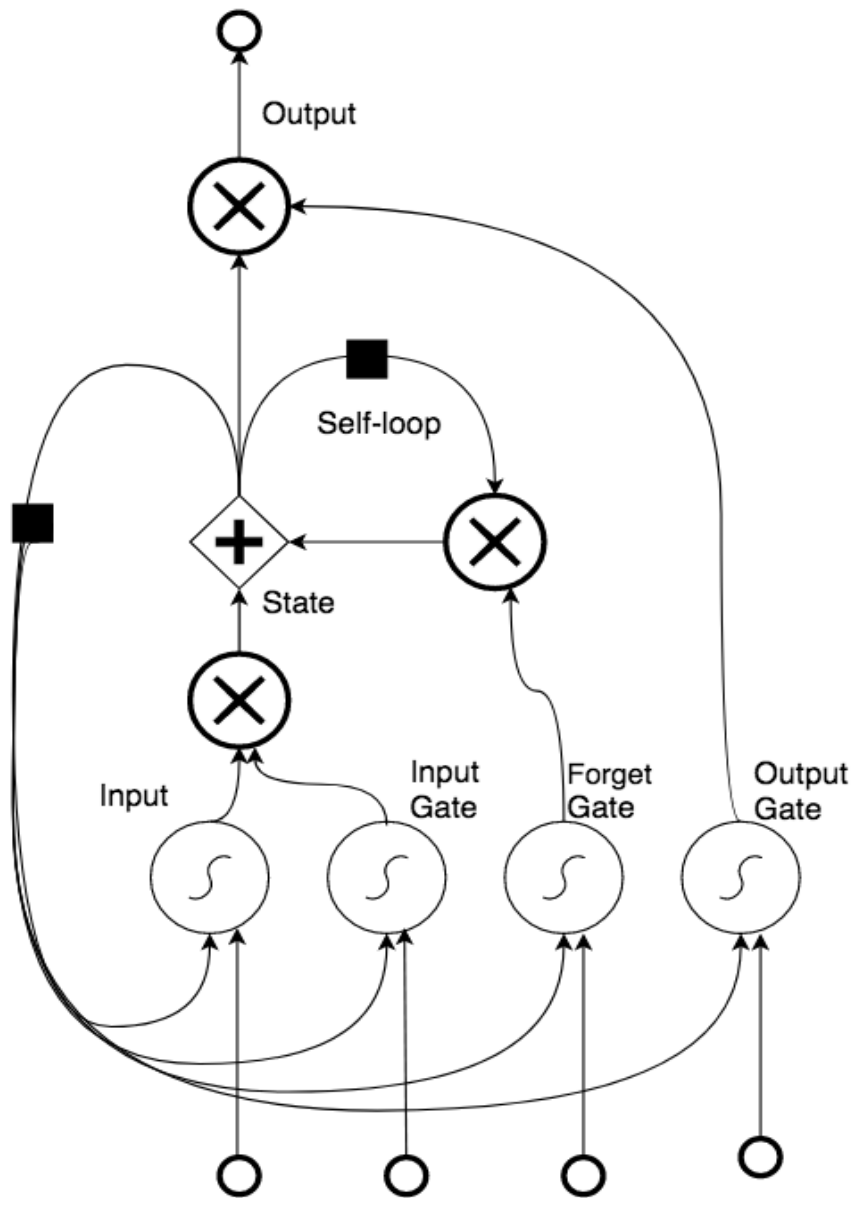

Figure 1 LSTM cell block diagram. The forget gate acts like "throttle" controlling the amplitude of the input, output, and self-loop, respectively.

### 3.4. Gated Recurrent Units (GRU) Cell

GRU cell is a special kind of gating mechanism for the LSTM cell of an RNN. It has simpler structure and thus fewer parameters than the LSTM cell. As a simplified version of LSTM cell, the GRU cell was proposed by (Chung, et al., 2014), which has less gates and thus less weights to be tuned. There are only two kinds of gates for GRU: reset gate, r, and update gate, z. The reset gate determines how to combine new input with that of the previous memory in its cells. On the other hand, the update gate decides the portion of its previous memory does it retain or throw out. The equations for the GRU cell are similar to the LSTM:

$$z = \sigma_z(x_t U^z + h_{t-1} W^z)$$
$$r = \sigma_r(x_t U^r + h_{t-1} W^r)$$
$$h = tanh(x_t U^h + h_{t-1} W^h)$$
$$h_t = (1-z)*h + z*h_{t-1}$$

Since there are only 2 gates, the important difference between a GRU versus an LSTM is that GRU does not possess an internal memory ($c_t$). The reset gate is directly applied to the previous hidden state ($h_{t-1}$). Thus, it can train faster since there are only few parameters in the U and W matrices and hence maybe useful when not much training data is available.

### 3.5. RNN Training Process with Long Term Dependencies

The *Early Stopping* technique is used for the training process. the number of iterations is the count of looping through the whole training set, and the Mean Square Error (MSE) of the training set and validation set is estimated for each loop. The MSE of the validation data set is used as a reference to meet certain criterion about the MSE of the system state estimation. For each time step, the weights in the RNN model is updated by the *Back Propagation Through Time (BPTT)*, which is a gradient-based technique to train the RNN. Specifically, the Adaptive Moment Estimation (Adam), having the best performance is used for the BPTT, (Kingma, et al. 2014). Since the BPTT multiplies gradient over a large number of time steps, the resulted gradient either get vanished or explode over multiple layers. The recurrent net tends to induce LSTM cell and GRU cell to avoid the gradient vanishing occurrence in simple RNN. On the other hand, the exploded gradient can lead to the training process to be unstable, leading to the algorithm to find a non-optimal solution. This is solved by *clipping the gradient* to a determined maximum value. As stated previously, the RNN for System Transition and System Observation are trained independently. The RNN cells (Elman RNN cell), LSTM cell, and GRU cell will be tested and compared on the performance index of their online estimation. The Mean Square Error (MSE) will be applied as a criterion for these three different cells. Moreover, as model-free methods, there is no assumption that there is any certain forms of relationship between the variables at time domain, regardless whether they are linear or nonlinear.

### 3.6. Diagnosis and Prognosis with RNN

After the three kinds of RNN model of System Transition and System Observation is trained, they can be further applied for diagnosis and prognosis of the system state, by inputting the indirect measurement in real time. Our proposed method is to realize online diagnosis, by taking advantage of the indirect measurement from past and current states: $p(y_t, \theta|x^t)$, where $x^t = (x_1, \ldots, x_t)$. For prognosis, following the method of the generative recurrent neural networks (Graves, 2013), the produced values from output nodes and the hidden nodes of the RNN of the System transition is inputted into the input node and hidden node of the model itself. This process is repeated, and the number of the repetitions is equal to the predicted number of time step horizons. Lastly, for Remaining Useful Life (RUL) prediction, the criterion of the degradation is set to certain value, for example 0.3 mm for the tool wear. By apply the generative model of RNN for system transition, until the predicted degradation variable reaches the criterion, the number of time step it passes through can be the predicted RUL, which in our case, will be the number of time steps it will take to reach the tool wear criteria.

## 4. Experimental Evaluation

### 4.1. Setup

In order to evaluate the performance of the LSTM based RNN method, experiments were conducted to obtain actual tool insert wear measurements using the HAAS VF2 CNC mill machine under dry milling machining conditions. A two-flute indexable milling tool with diameter 12.7 mm, Sandvik Coromill 390 (RA390-013O13-07L) installed with one uncoated insert, Sandvik Coromill 390 Insert (390R-070204M-PM) was used in the cutting process. The workpiece material utilized is Steel 4142 (cold rolled, 40-45HRC, ). Each single cutting pass was 37.0 mm long, with the spindle speed at 500 rpm, leading to a surface speed of 19.81m/min and chip load at cutting to be 0.05mm. The radial depth of cut is 6.5 mm and axial depth at 0.4 mm. Feed rate is 50 mm/min. A 3-axis vibration sensor (Kistler accelerometers 8762A10) was installed on the fixture attached to the workpiece (Figure 2). Sensor data was sampled at 1652Khz and its RMS signal formed the online indirect measurement. No force dynamometers were used in this study to emulate minimally intrusive sensing. After every single pass of cut along the length of the workpiece, offline direct measurements were taken to measure actual tool wear. The machining process was repeated for two experimental runs. The RMS of the vibrations signals for the two repeated runs are shown in Figure 3. The vibration signal curves do not monotonically increase with the tool wear. The signal curves show that it is critical to account for the time dependency for any diagnosis and prognosis of cutting tool wear.

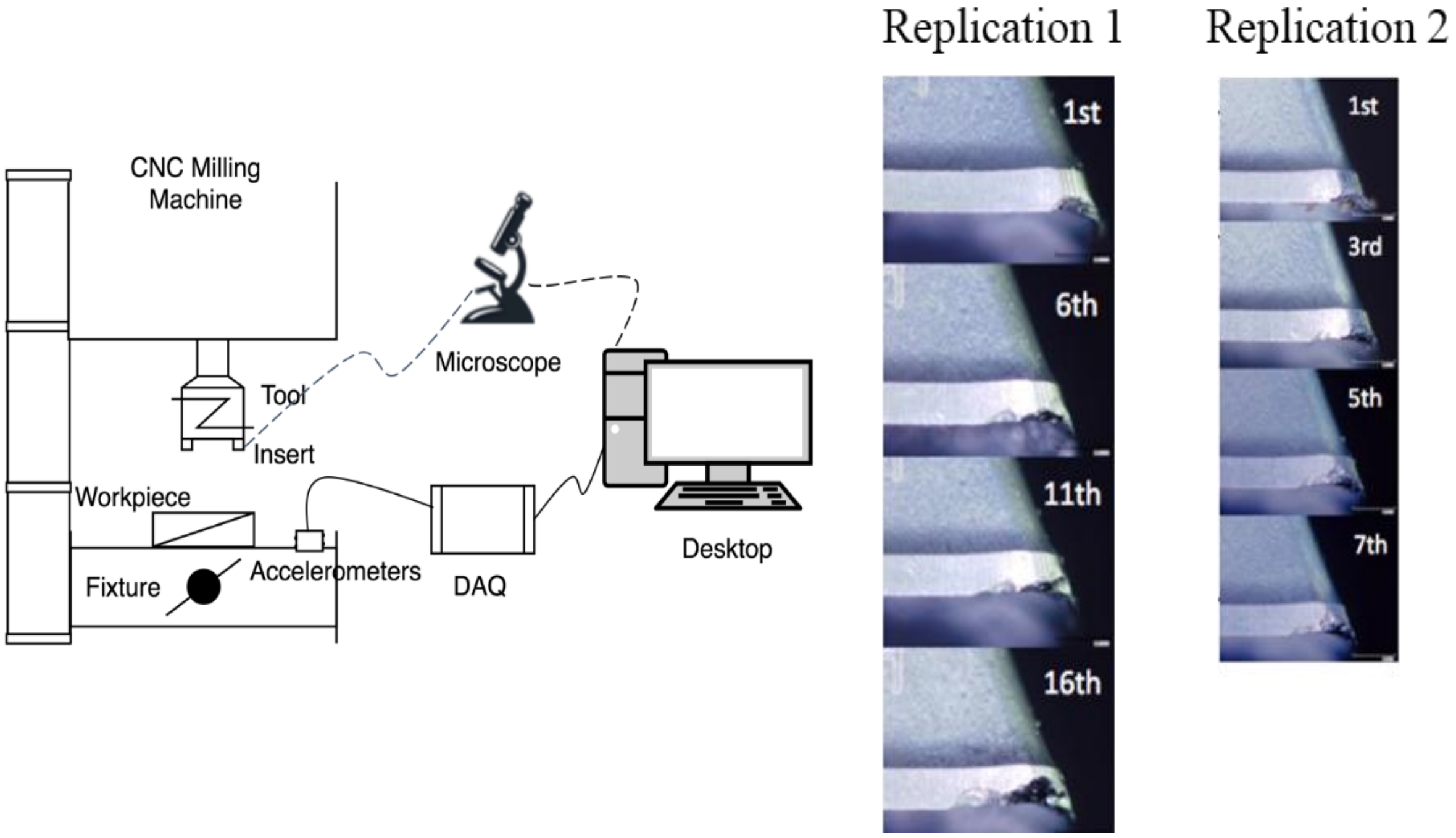

Figure 2 Schematic diagram of experimental setup with micrographs of the cutting tool surface at various cut path numbers.

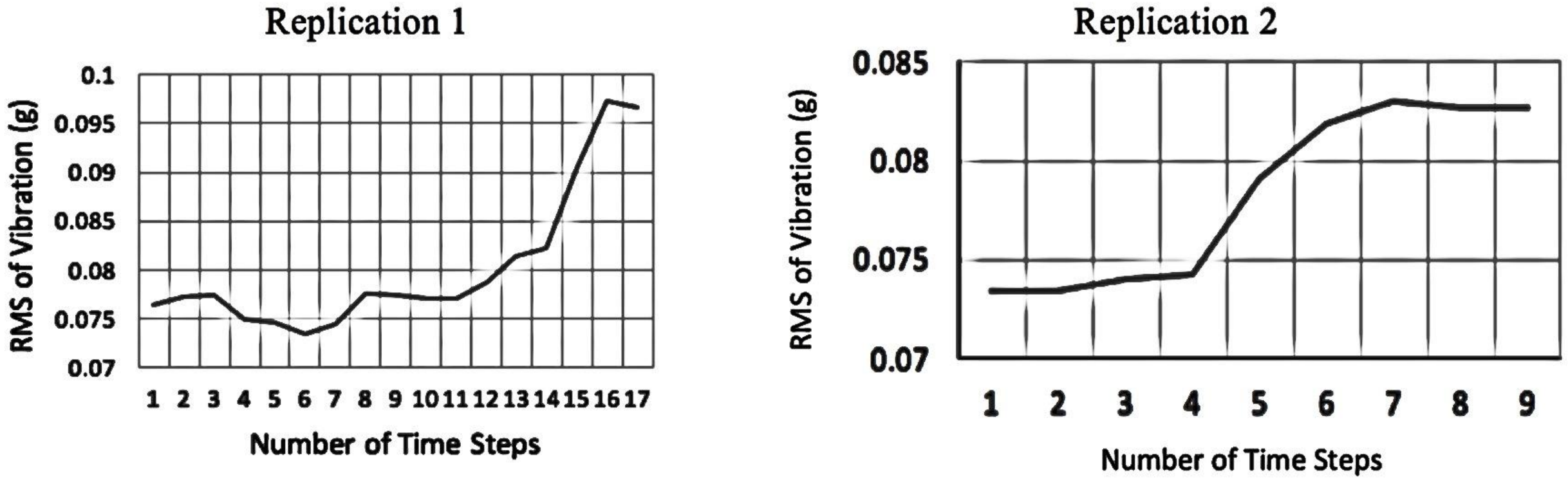

Figure 3 Vibration Index for two replications of cutting the same materia under similar machining conditions. The only difference is the each replication was carried out by a brand new cutting tool insert but of the same specifications.

### 4.2. Model Training

Early stopping is applied to loop through the training set for a number of iterations for the three kinds of RNN model for the system transition and observation functions, respectively. The mean square error (MSE) of the validation set is evaluated for each iteration. The whole dataset is divided into three parts, the first one is the training set (7 pairwise sequences), the second one is the validation set (6 pairwise sequences), and the last one is the test set (2 pairwise sequences). For each time step, the weights in the RNN model (U and W) are updated by the backward propagation (BPTT). For each sequence, the values of hidden nodes is reset to zero. The RNN for System Transition and System Observation are also trained independently.

During the training process, the reduction rate of the average MSE decreases as the training proceeds. The iteration number is set as 14, given the computational capacity and average MSE is good enough for estimation. Comparing the MSE values for the System Transition function, the results show that the average MSE of the LSTM is lower than the other two - Elman RNN and GRU. The Standard deviation (Std.) also shows that the stability of the LSTM to have relative good performance for both system transition and observation (see Table 1).

Table 1 Compare MSE of validation set for RNN, LSTM, and GRU for System Transition and Observation

| **RNN Cell Type** | **System Transition** | | **System Observation** | |
|---|---|---|---|---|
| | **Avg. of MSE ($mm^2$)** | **Std. of MSE ($mm^2$)** | **Avg. of MSE ($mm^2$)** | **Std. of MSE ($mm^2$)** |
| Elman RNN | 0.0027180 | 0.0007639 | 0.0038187 | 0.0016212 |
| LSTM | 0.0015628 | 0.0007708 | 0.0037855 | 0.0003046 |
| GRU | 0.0030236 | 0.0010090 | 0.0038729 | 0.0010628 |

All the RNNs have one hidden layer with 5 hidden nodes. The MSE of validation set of trained LSTM for System Transition function is 0.0015628 $mm^2$, and the MSE of validation set of applied trained LSTM of the System Observation function is 0.0037855 $mm^2$. Since there are more parameters in the LSTM cell than the other two, Elman RNN and GRU, it shows its flexibility to fit the data and exhibit more adaptive properties to construct the complexity of the time dependencies, although more computational resources maybe needed for determining the parameters in the model.

## 5. Results

### 5.1. Online Diagnosis

By applying the LSTM for system observation, given the indirect measurement of the current and past time step, the RMS of the vibration signal, the model is able to estimate the tool wear at current time step.

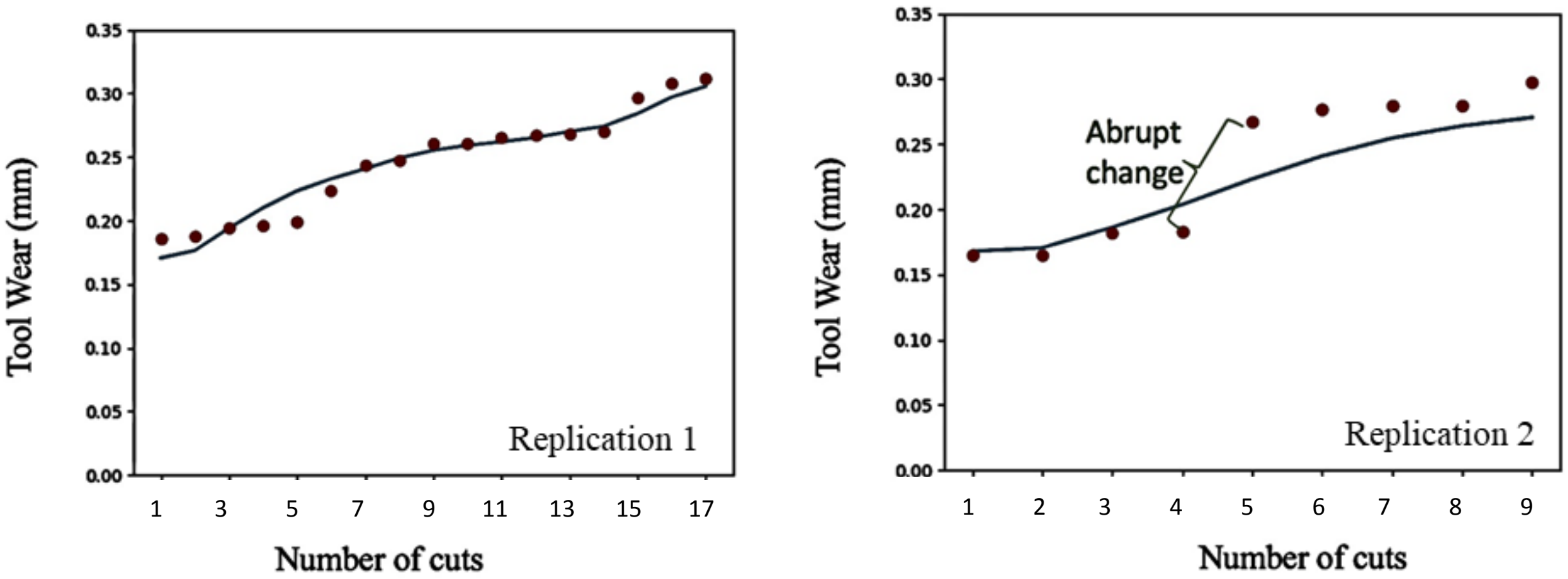

Figure 4 Online Tool Wear Diagnosis with indirect measurement. The dots indicate the actual measured tool wear; the solid lines are the estimated tool wear. A) Replication 1; B) Replication 2

For Replication 1 (Figure 4A), the estimated tool wear closely fit the real tool wear, and thus the curve shape of the estimation appear more like a bending line, rather than a smooth curve, thanks to the online estimation of the tool wear which is adjusted based on the indirect measurement. For the Replication 2 (Figure 4B), there is an abrupt tool wear change between the 3$^{rd}$ to the 4$^{th}$ cut. The estimated tool wear adjusts for approaching the real tool wear, with certain time lag. This illustrates that the method is robust in certain extent on tolerating abrupt change.

**Online One-Step and Two Step Ahead Prediction**

For one-step ahead prediction, the estimated tool wear at current time step is input into the input node of the LSTM for the System transition function. It assumes that the tool wear at time step 0 is equal to 0.0mm ($y_{t=0}$ =0.0 mm). The predicted tool wear at next time step t+1 is the value from the output node of the RNN for the System transition, given the hidden node and estimated tool wear at current time step t. This means that the tool wear at next time step is predicted given the current and past tool wear estimation as given by the system observation function (Figure 5). Similar to online one-step ahead prediction, the tool wear at two time-step ahead is predicted given the next, current and past tool wear estimates by the system transition function and the system observation function and (Figure 6). The predicted tool wear for one-step and two-step ahead appear smoother than online estimated ones, and do not closely fit the tool wear. They provide the prediction of the tool wear and show the trend of the tool wear. When the time step proceeds to the predicted cutting step, the estimation is updated according to the online indirect measurement.

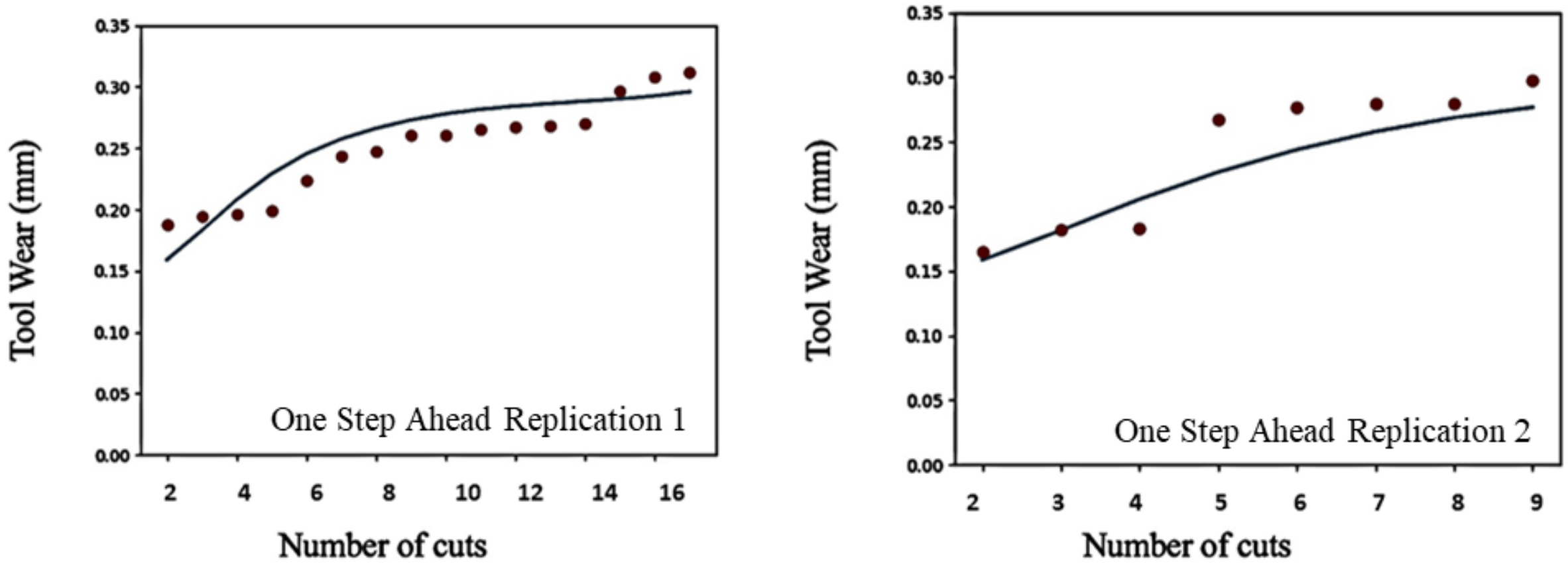

Figure 5 Online Tool Wear One-Step Ahead Prediction with indirect measurement. The solid circles indicate the actual tool wear; the solid lines are the predicted tool wear.

The proposed method is robust to certain abrupt change in the system state, the tool wear. Arbitrary number of time step ahead prediction can be realized by the proposed method. Similar to online two-step ahead prediction, the tool wear at three or multiple time step ahead can be predicted given the two step ahead or multiple time minus one step ahead, current and past tool wear estimation by the system transition function and system observation function respectively.

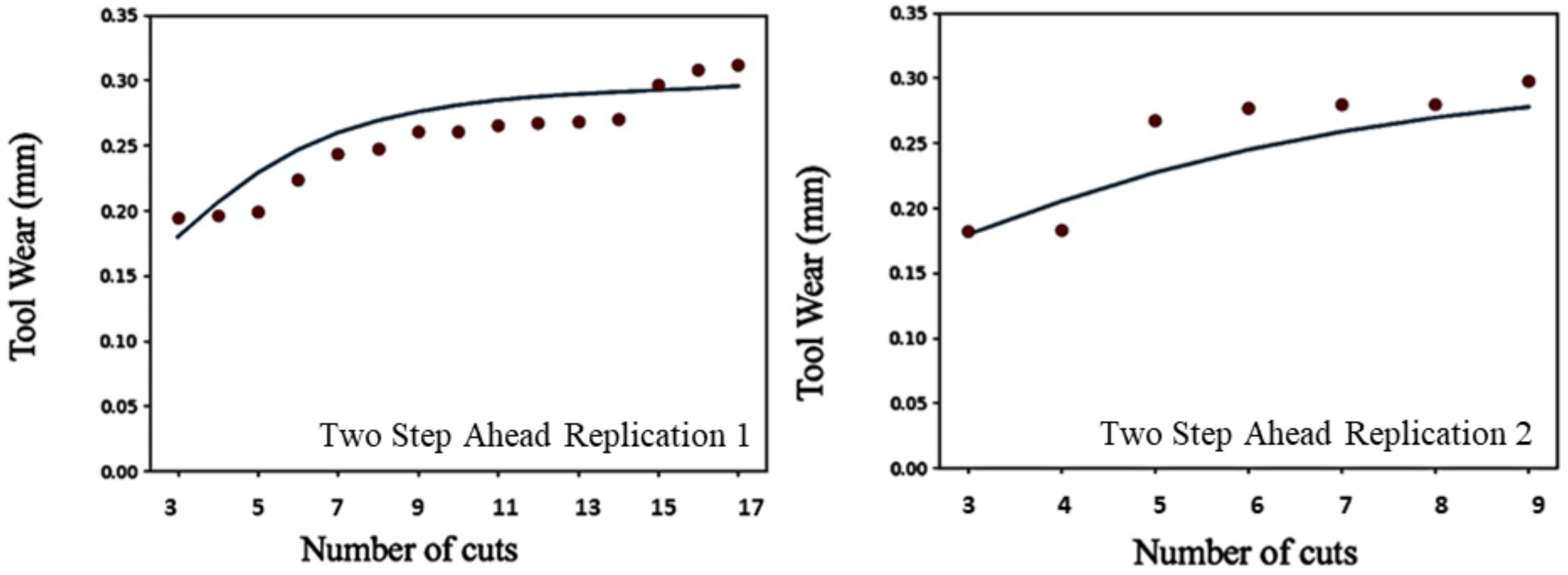

Figure 6 Online Tool Wear Two-Step Ahead Prediction with indirect measurement. The circles indicate the actual tool wear amount; the solid lines are the predicted tool wear.

### 5.2. Remaining Useful Life (RUL) Prediction

At each time step, by looping the RNN for system transition, which means if we input the values of the hidden and output nodes of RNN for system transition of the last time step into the hidden nodes and input nodes of the next time step, and repeat the process, until the value of the output nodes reaches the criterion for the tool wear set to be at 0.3 mm. Both of the RUL predictions start at 12 cuts (Figure 7). For replication 1 and 2, the predicted RUL is updated according to the online indirect measurement, so that the predicted RUL can be adjusted to approach the real RUL.

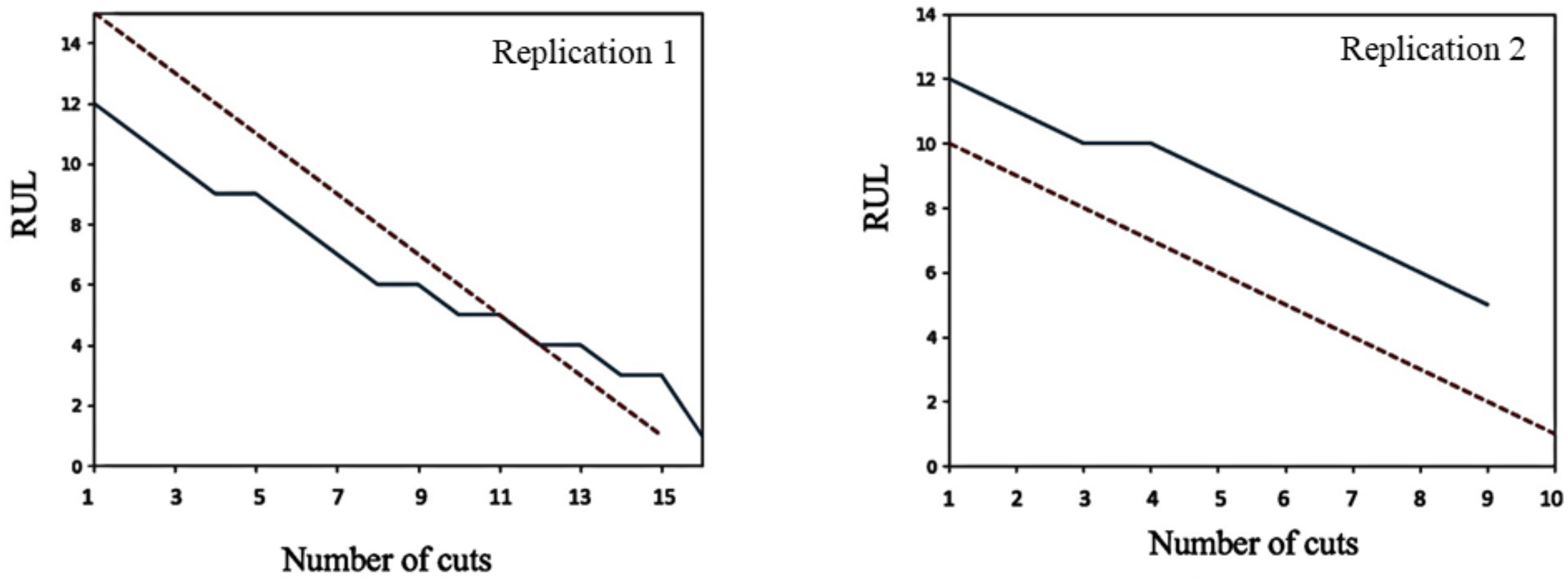

Figure 7 Online RUL Prediction. The dashed lines indicate the actual RUL; the solid lines are the predicted RUL.

## 6. Discussion

LSTM and GRU, as gated RNN, can capture the long term time dependencies as a model for the system transition and observation functions, compared to the simple RNN and Elman RNN. By taking advantage of the trained RNN system observation model, the tool wear can be estimated from currently observed indirect measurement and past indirect measurements. Also, for tool wear prediction, one time step ahead tool wear prediction can be realized by input into RNN system transition function, the estimated current and past tool wear by the RNN system observation function. For two step ahead tool wear prediction, the one step ahead tool wear prediction by the RNN system transition function and all the estimated tool wear by the RNN system observation function act as inputs into the RNN system transition model, thus, resulting in the two step ahead tool wear prediction. Similarly, the multiple time step ahead prediction can also be estimated. This is attributed to the design of the generative RNN, in which our proposed model can predict arbitrary number of time step ahead tool wear degradation in the future.

Two replications of the tool were tested to demonstrate and evaluate the proposed model approach. It is to be noted that despite the same cutting conditions, the tool wear diagnosis and prognosis is quite different for the two replications. This is typical of all cutting tools since they can wear down at varying rates and patterns. We are able to predict the tool wear amounts on the fly, which confirms that the necessary diagnosis and prognosis methods can be adapted online. Furthermore, the RUL prediction can also be predicted until the tool wear reaches the set criterion.

In this study, the indirect measurement is the signal obtained from the vibration generated during the cutting process. As a black box method, the Neural Network model does not explicitly show the formula of the system observation and system transition function as seen in the Particle Learning method proposed by Zhang, et al, (2017) and any other analytic models typically used in cutting tool wear estimation. The LSTM based RNN approach can increase the computation cost, and the challenge here is to decide the complexity of the network, number of layers and number of hidden nodes, given the reward is the flexibility of the fitting any form of relationship. Additionally, the system transition function also needs to be trained beforehand before applying the prognosis process. This method can have desirable diagnosis and prognosis result especially if the tool wear gradually occurs but also can tolerate certain abrupt tool wear change. However, for different material and different machining conditions, new data should be collected to train a new model.

The LSTM based RNN can make prediction more superior than the feed forward neural network. Second, the approach is also better than the time-delay neural network, which can only predict limited time step ahead of machine states. Third, for the RNN with only time-delayed form, it cannot realize the arbitrary number of time step ahead prediction, not to mention not being able to predict RUL. Fourth, Malhotra, et al. (2016) used the RUL as target variable and sensor signal to be modeled with the RNN. However, the RUL does not reflect any physical characterization, so that there is no direct physical connection between the sensor signal and RUL. Also, the approach of using RUL as target variable cannot realize machine state diagnosis and prognosis. Compared to these methods, the approach of modeling the system transition function and system observation function with the RNN renders the online estimation considering the time dependencies become natural. Taking the schema of the generative RNN, arbitrary time step ahead prediction get realized by combining the system transition function and system observation function modeled by RNNs. Further, by applying advanced form RNN cell, the LSTM, GRU, the long term time dependencies can be captured.

The neural network model is a more flexible method than the analytic model, which do not need physical formula within the model. The proposed approach by modeling the system transition and system observation with long term time dependency helps realize the online estimation of the system state and predicts the arbitrary future horizon of the system state, by the given current and past indirect measurements. The application of this method can be extended into any critical components of the physical machine, such as bearings, gears, engine, ball screws, grinding wheels, nano-machining processes and bioreactors, in which only indirect measurements are accessible in real time and there is a need to estimate and predict the system state.

There are several possible ways to improve the accuracy of the prediction. First one is to loop through more number of iterations of the training set to train the model; second one is to use more hidden nodes or more layers of hidden nodes and even combining the *dropout* technique (Srivastava, et al., 2014) within the neural network model; third one is to collect more data to train the model. With more data to train the model, and more available data for the validation, the model can get more robust and have more stable performance in diagnosis and prognosis. Moreover, one can also use the n-fold or leave-one-out cross validation method to get stable evaluation of the MSE of both the RNN for system transition and RNN for system observation. Incorporating more time domain and frequency domain feature of the signal into the inputs of the model, so that feature engineering work should be implemented.

## 7. Conclusion

The paper proposes a framework for the machine state diagnosis and prognosis using RNN models for system transition and observation function via indirect measurements. The LSTM and GRU, which can capture the long term dependencies is used to model those two functions. The validation set was used to evaluate the estimation accuracy by the MSE. Two replications of sequence of the indirect measurement, as test set, is used by the proposed model, to demonstrate the online estimation, one and two time-step prediction, and RUL prediction. The results were compared with the real tool wear and showed good fit. RUL prediction was also compared with the real RUL. The validation set shows acceptable MSE for both system observation and transition functions. After comparison with the real values, the results show good performance and strong possibility for practical usage. First, as model-free methods, the Neural Network requires no analytic formula in modeling. Second, the LSTM cell modeled system transition and system observation functions, which capture the long term temporal dependencies, has better performance than the simple RNN to model them. The arbitrary horizon of prediction can be realized by taking advantage of the modeled system transition function and the generative model, and thus the RUL prediction. It is very possible to extend the proposed methods for diagnosis and prognosis of system state by indirect measurement for other physical entities, such as gears, bearings, ball screw, motors, etc.

**Acknowledgement**

We acknowledge the financial support from the US Department of Defense (DoD) - DMDII 15-16-08 towards collection of machine data.

| RNN Cell Type | System Transition | | System Observation | |
|---|---|---|---|---|
| | Avg. of MSE ($mm^2$) | Std. of MSE ($mm^2$) | Avg. of MSE ($mm^2$) | Std. of MSE ($mm^2$) |
| Elman RNN | 0.0027180 | 0.0007639 | 0.0038187 | 0.0016212 |
| LSTM | 0.0015628 | 0.0007708 | 0.0037855 | 0.0003046 |
| GRU | 0.0030236 | 0.0010090 | 0.0038729 | 0.0010628 |

Figure(s)
Click here to download high resolution image

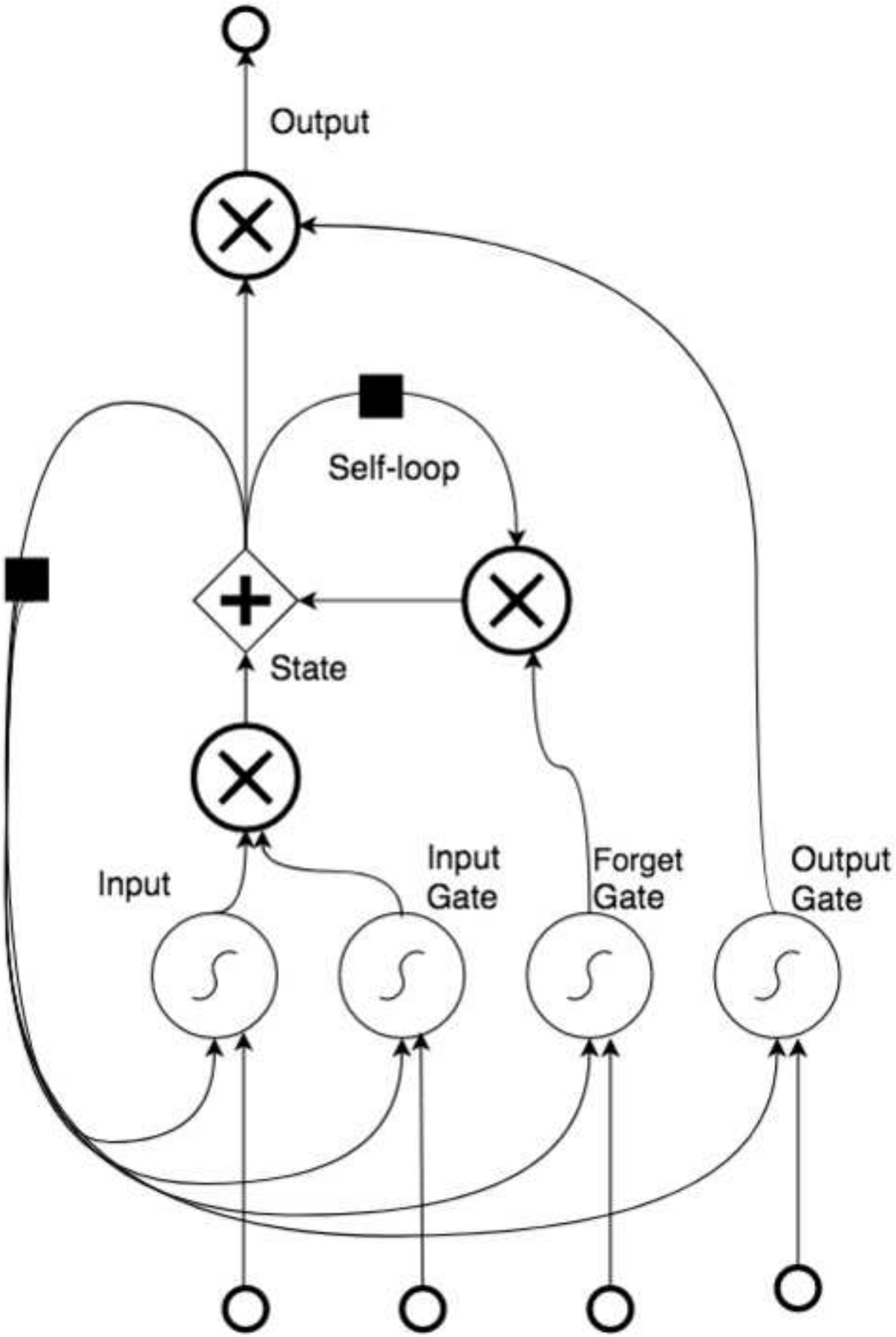

Figure(s)

Click here to download high resolution image

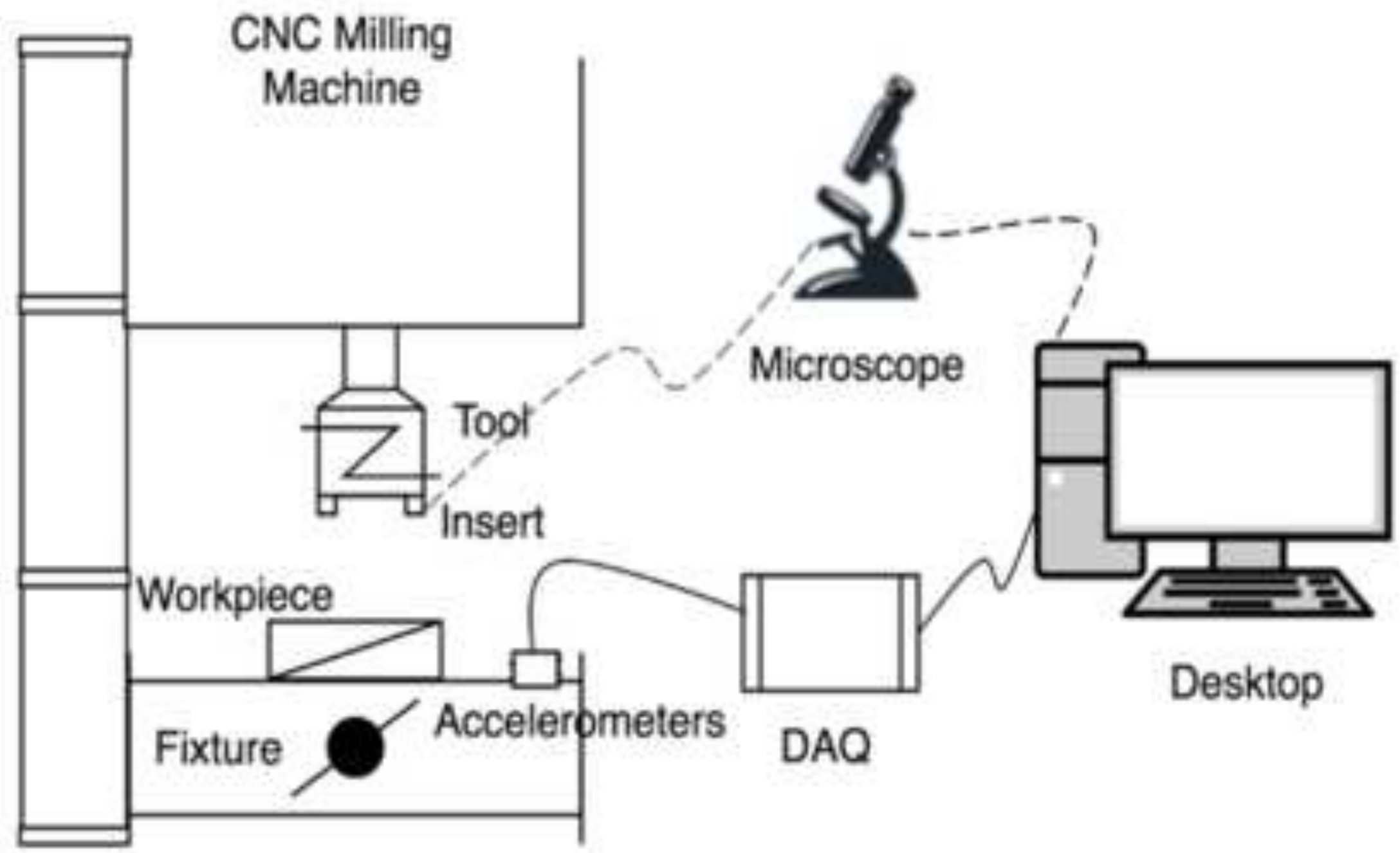

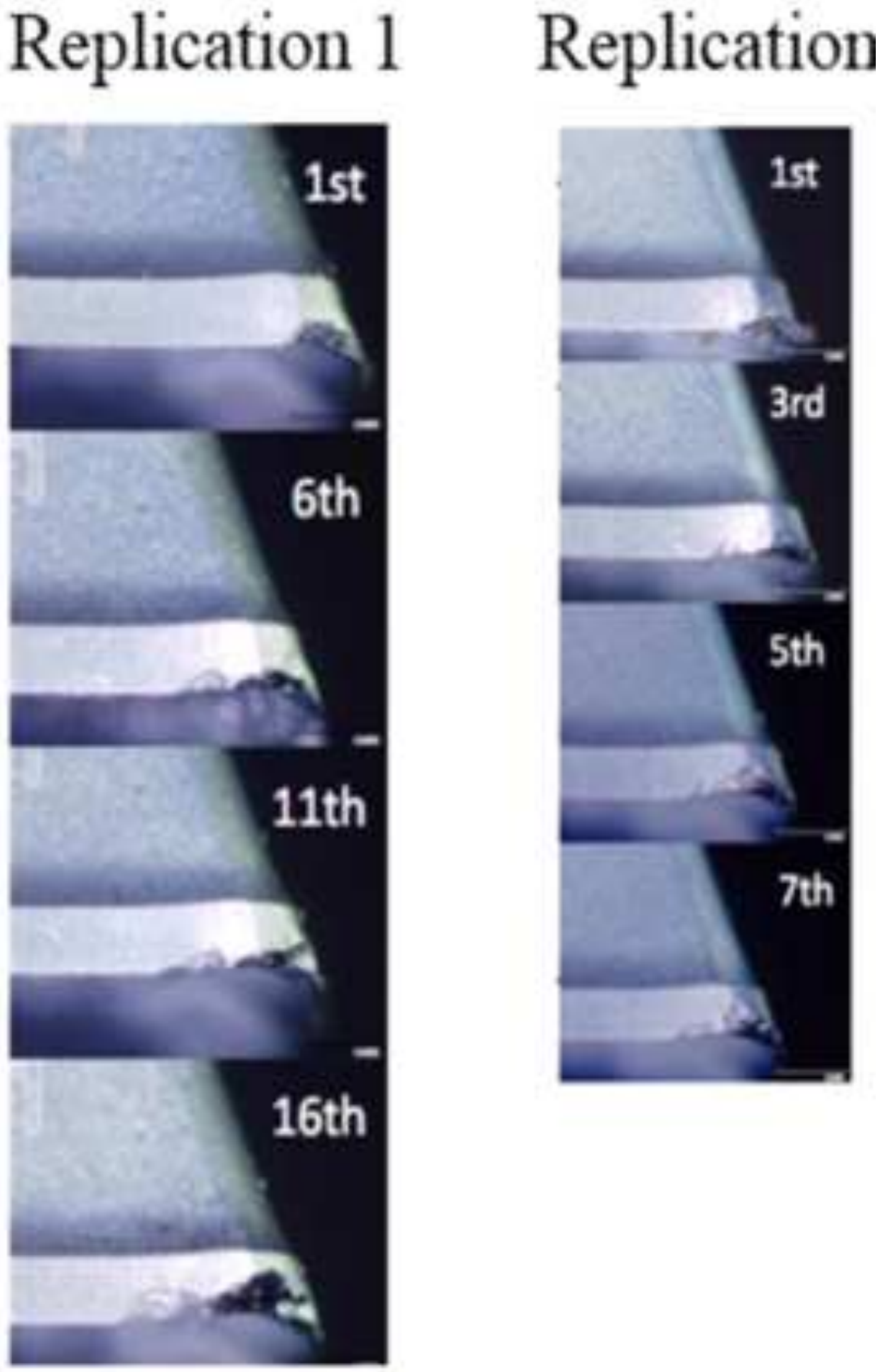

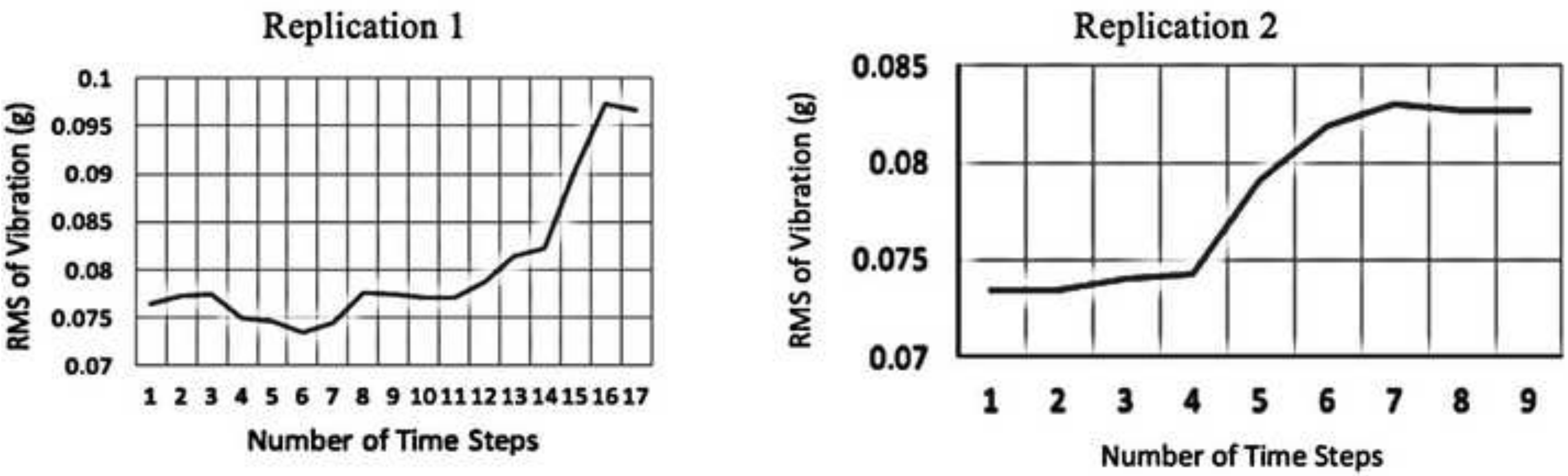
Replication 1
RMS of Vibration (g)
0.1
0.095
0.09
0.085
0.08
0.075
0.07
1 2 3 4 5 6 7 8 9 10 11 12 13 14 15 16 17
Number of Time Steps
Replication 2
RMS of Vibration (g)
0.085
0.08
0.075
0.07
1 2 3 4 5 6 7 8 9
Number of Time Steps

Figure(s)
Click here to download high resolution image

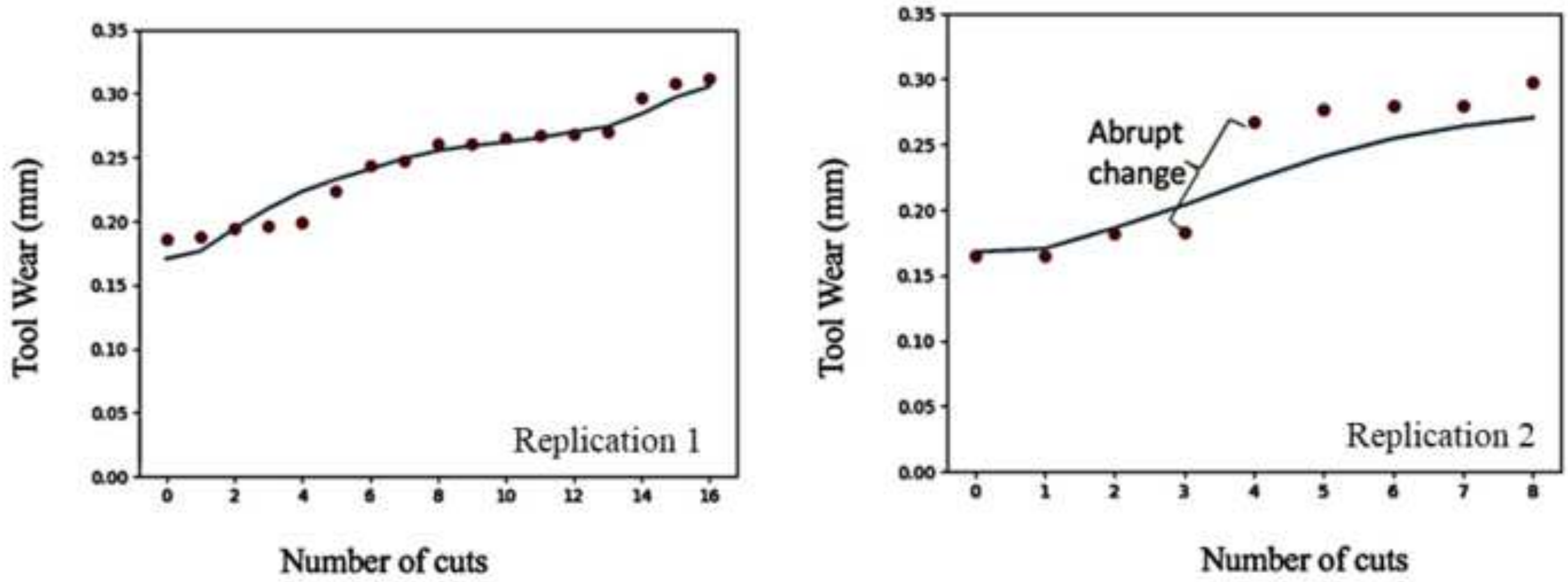

Figure(s)
Click here to download high resolution image

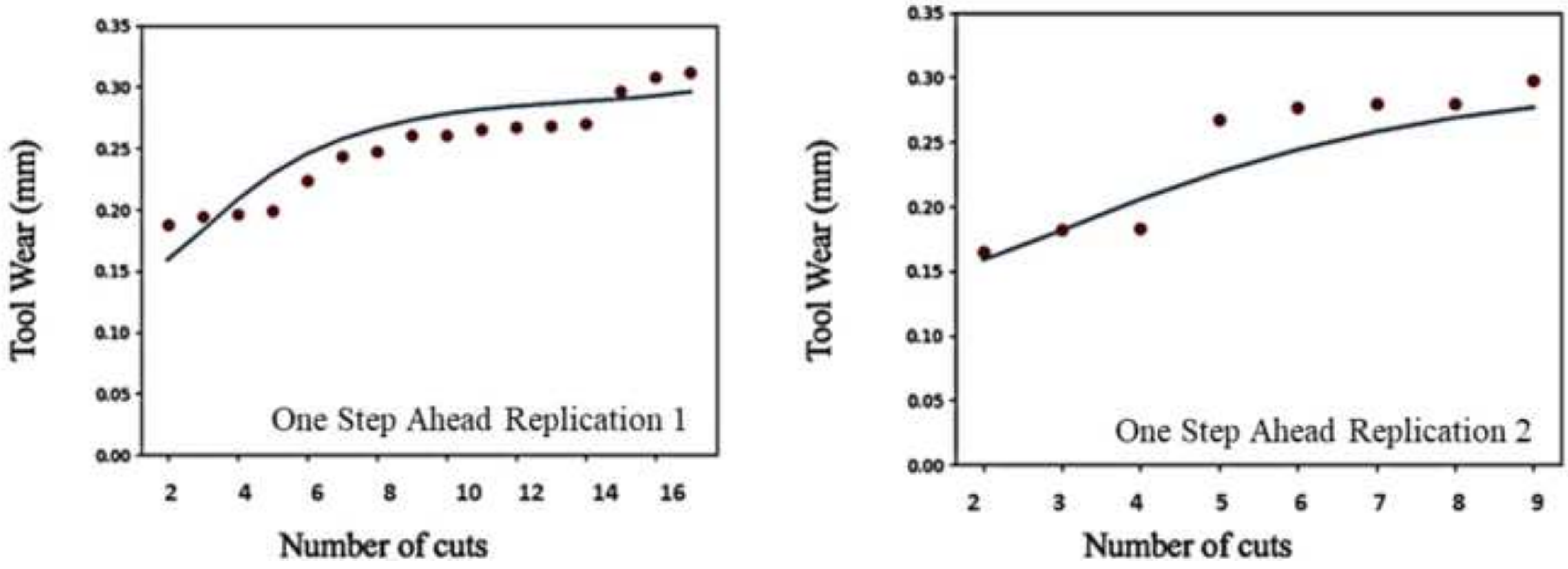

Figure(s)
Click here to download high resolution image

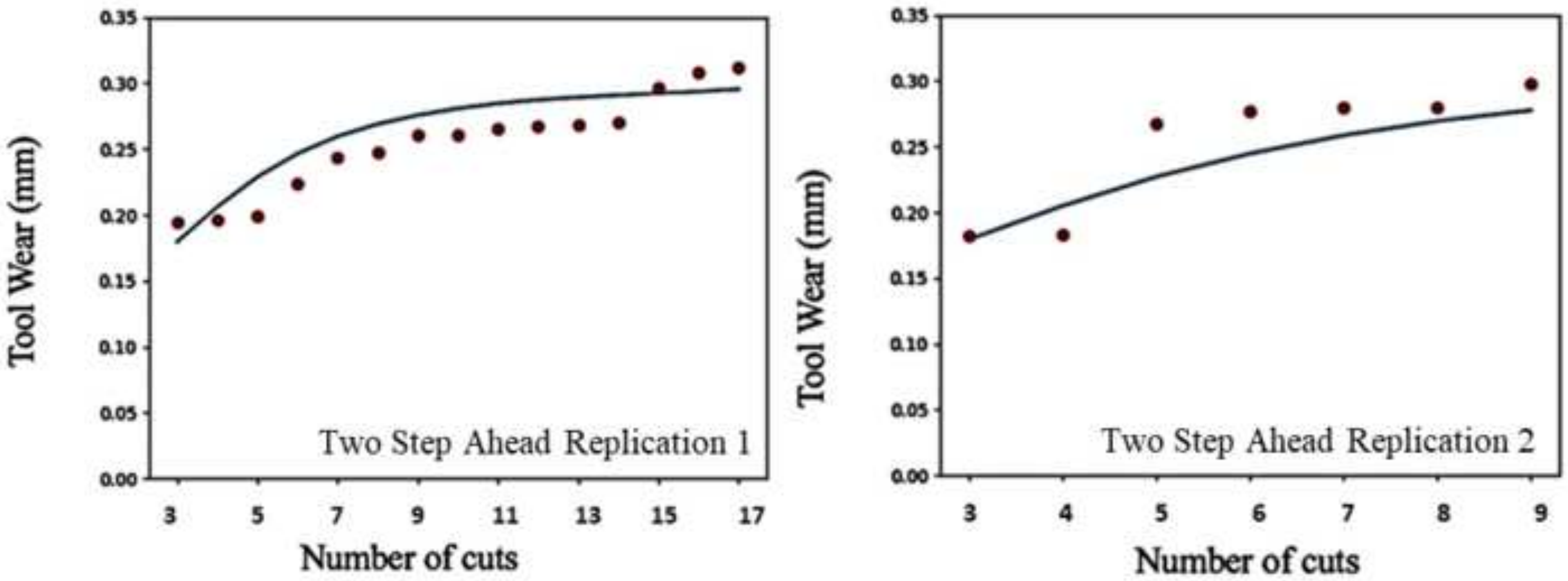

Figure(s)
Click here to download high resolution image

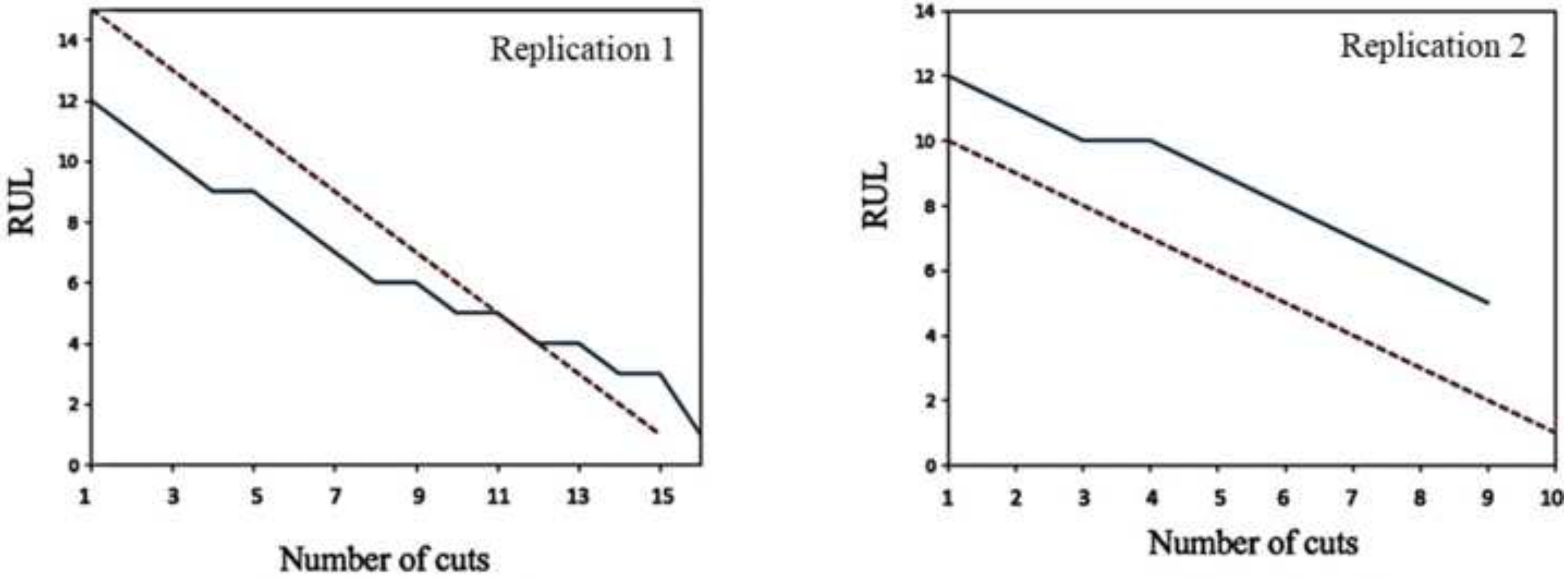